\documentstyle[aps,prl]{revtex}
\draft

\begin{document}

\title{An oriented process induced by dynamically regulated energy barriers}
\author{Naoko Nakagawa}
\address{
         Department of Mathematical Sciences,
        Ibaraki University, Mito, Ibaraki 310-8512, Japan\\
        Laboratoire de Physique, ENS-Lyon, 46 all\'ee d'Italie,
        69364 Lyon Cedex 07, France
}

\author{Teruhisa S. Komatsu}
\address{Department of Physics, Gakushuin University,
        Mejiro, Toshima-ku, Tokyo 171-8588, Japan
}

\date{14 April, 2004}

\maketitle

\begin{abstract}
A novel mechanism for the appearance of
 oriented processes is investigated with a flexible dynamical system
overcoming barriers.
Under non-equilibrium condition with external driving,
 reaction paths deviate from that at equilibrium
 with an accompanying violation of symmetry between the forward 
and the reverse paths.
Although we never introduce any external switching of potentials
 to generate the oriented processes,
 multi-dimensional flexible dynamics promote the oriented processes
 through this symmetry violation.
Along the reaction paths, 
bottleneck points are proposed as a rate-controlling factor, 
which determine the {\it direction-dependent activation energies} satisfying
Arrhenius-like law for the rate constants.
In comparison, in stiff systems, the oriented process is suggested 
 to appear in different manner from this scenario.
\end{abstract}


\section{Introduction}

In thermal equilibrium, free energy difference is considered 
 as the single driving force to generate oriented processes. 
On the other hand, far from equilibrium, biomolecular processes
 are often said to be ''efficiently'' facilitated.
Then other mechanisms to generate oriented processes
 may be explored in non-equilibrium situation.
One possible example may be clearly found in a biomotor 
 showing directional stepping motion along a filament \cite{Review-motor}. 
The stepping of the motor is a kind of {\it reaction} shifting
 the binding site from one to the neighbors. 
Since its position does not contribute to the free energy,
 the directionality is not explained by the free energy difference 
 between before and after this reaction.
As ATP-hydrolysis drives the motor far from equilibrium,
 the stepping reaction may not be explained by the rules that govern 
  equilibrium oriented reaction. 
Some experimental works \cite{Nishiyama_Higuchi,Taniguchi_Nishiyama}
suggest that the effect of ATP-hydrolysis for the stepping reaction
is to generate the direction-dependent activation energies.
Such dependency certainly induces oriented process of the motor, 
but the question is then to understand how such a dependency can appear 
in non-equilibrium. 

The generation of oriented processes is a problem of general importance
in biological systems.
Protein folding is facilitated far from equilibrium
in some chaperonins with ATP-hydrolysis \cite{Review-Chaperonin}.
Although the folding processes are originally oriented by free energy difference, 
the possible facilitation mechanism would act additively to accelerate the processes.
Quick conformational changes of proteins after excitations \cite{Champion}
might be facilitated in a similar way. 
Keeping several possible examples in mind, we want to propose 
mechanisms which lead to oriented processes.
In this paper, we never introduce any external switching of potentials to 
generate the oriented processes, but, 
instead we show that 
multi-dimensional dynamics can
 promote an oriented process provided the system is far from equilibrium.

\section{Model}

Many chemical reactions 
can be modeled by barrier overcoming process.
The shift of a {\it particle} in one dimensional periodic potential,
a standard theoretical frame for particle current or diffusion \cite{Risken},
gives a model to consider steady state for barrier overcoming process.
Brownian ratchet models \cite{Review-Ratchet} possess similar structure, 
and the stepping process of biomotor systems, too. 
While referring to these preceding studies, we introduce a new feature 
that the periodic structure is not fixed but fluctuates dynamically.
Consider a {\it rail} on which stable binding sites of the particle
 are arranged periodically.
When the rail is modeled as a {\em flexible chain},
the particle senses the interaction potential 
 modulated by the temporal conformations of the rail, 
 i.e. modulated by the dynamics of the system itself \cite{Nakagawa_Kaneko4,Nakagawa_Kaneko6}.

Non-equilibrium situation is introduced by 
two heat baths,
 one of which with temperature $T_1$ is attached to the rail
 and the other with $T_2$ to the particle.
Although this setup seems a little bit artificial, it has the clear advantage that
a non-equilibrium steady state of the system is achieved.
The resulting system may resemble to Feynmann's ratchet heat engine \cite{Feynmann}, 
 but in the present model the internal degrees of freedom of the rail
are explicitly considered.

A schematic picture of the model is shown in Fig.\ref{fig:model}.
Independent variables are
 the position $x_h$ of the particle and those of the lattice sites $\{x_i\}$
for the 1D chain,
 where $i$ is the site index.
The particle and each lattice site interact through the potential $V(x_h-x_i)$,
 which has asymmetric shape with one minimum as shown in Fig.\ref{fig:model}.
The particle senses the summation of these potentials,
$V(x_h,\{x_i\})\equiv\sum_i V(x_h-x_i)$.
The time development is described by Langevin equations,
\begin{eqnarray}
\ddot x_i &=& -\gamma \dot x_i +\sqrt{2\gamma k_B T_1}\xi_i(t)
-K_c\left\{ (x_i-il)+(x_i-\frac{x_{i-1}+x_{i+1}}{2})\right\}
-\frac{\partial V(x_h-x_i)}{\partial x_i}, \nonumber\\
\ddot x_h &=& -\gamma \dot x_h +\sqrt{2\gamma k_B T_2}\xi_h(t)
-\frac{\partial V(x_h, \{x_i\})}{\partial x_h}, 
\label{eqn:model}
\end{eqnarray}
driven by two heat baths ($T_2\geq T_1$). 
We adopt sufficiently small values of the temperatures $T_1$ and $T_2$
 comparing to the potential well depth of $V$.
$\gamma$ is a friction coefficient 
and $\xi_{\alpha}(t)$ represents Gaussian white noise.
We use the units as Boltzmann constant $k_B=1$.
Although Eqs.(\ref{eqn:model}) include inertia terms, 
 qualitatively the same results are obtained
 in overdamped cases without these terms.
$K_c$ is the spring constant for the chain with which 
each lattice site is connected to neighbors, and also to fixed ground,
with the natural interval length $l$.

Hereafter two typical values of stiffness $K_c$ are used for the demonstration:
 at $K_c=0.5$, the flexibility of the chain greatly affects dynamics of the system (flexible system),
while at $K_c=5.0$
the dynamics of the internal degrees of freedom for the chain are considerably suppressed
similarly to the cases of certain Brownian ratchets (stiff system).

\section{Rate constants defined from time sequence}

A typical time sequence of the particle position
is shown in Fig.{\ref{fig:sequence}}, in which
we can observe each jump (stepping process) of the particle.
As is seen there, 
the particle steps to both direction with equal probability when $T_1=T_2$, 
whereas oriented motion appears under the existence of temperature difference
($T_2>T_1$).
In the following, 'forward' direction is defined for each value of $K_c$ 
 as the direction of the oriented motion in $T_2>T_1$, 
 and 'backward' direction as the reverse direction.
The forward and the backward stepping processes
correspond to their reverse process each other.

Because most of the successive stepping events are sufficiently
 separated in time, each process could be regarded as a stochastic one.
Then, it is natural to consider the probability of the steppings in a steady driven state.
The long-term observation of the particle position gives
 an ensemble of data how long the particle has stayed at one lattice site
 after previous stepping event.
Consequently we have a ``population" of waiting states
 as a function of waiting time.
Time evolution of such population could be analogous to chemical kinetics,
 and a rate constant for the occurrence of a step is introduced as follows.

Let $N_s(\tau)$ denote the number of events 
that
 the particle stays at a lattice site without stepping during a time longer than $\tau$ and
 $N_F(\tau)$ ($N_B(\tau)$) denote the number of events that
 the particle has stepped forward (backward) within $\tau$.
The rate of the forward (backward) stepping $\kappa_F(\tau)$
 ($\kappa_B(\tau)$) is defined as
\begin{equation}
\frac{d N_F}{d\tau} = \kappa_F(\tau) N_s(\tau),\quad
\frac{d N_B}{d\tau} = \kappa_B(\tau) N_s(\tau),
\label{eqn:rate-eq}
\end{equation}
where ${d N_F}/{d\tau}$ (${d N_B}/{d\tau}$) is 
 the number of the forward (backward) stepping events per unit time.

In the inset of Fig.\ref{fig:sequence}, $\kappa_F(\tau)$ and $\kappa_B(\tau)$ are
 shown for the case of $K_c=0.5$ for $T_2>T_1$.
It is found that
 these rates are almost constant for sufficiently large $\tau$.
For all cases examined with changing 
$T_1$, $T_2$ and $K_c$, 
converged values for
 $\kappa_F(\tau)$ and $\kappa_B(\tau)$ were obtained,
 which 
 implies the process can be regarded as
 an independent stochastic process for long $\tau$.
In the following, we let $\kappa_F$ and $\kappa_B$,
 which referred to the rate constants for ``stepping reaction'',
 denote the constant values of $\kappa_F(\tau)$ and $\kappa_B(\tau)$
 for large $\tau$. 

From the long time average of the stepping motion,
 the current and the diffusion coefficient of the particle
 are defined as
$
J \equiv \lim_{t\rightarrow\infty} {\langle x_h(t)-x_h(0) \rangle}/{t}$, 
$
D \equiv \lim_{t\rightarrow\infty} {\langle (x_h(t)-x_h(0))^2 \rangle}/{2t}.
$
For all the cases examined numerically, 
$J$ and $D$ converged to finite values.
In thermal equilibrium ($T_1=T_2$), the particle simply diffuses along the chain
 without drifting ($J=0$) all over the temperature range, 
while $J\neq 0$ in non-equilibrium ($T_2>T_1$). 
Interestingly, the direction of the drift in the flexible system ($K_c=0.5$)
 is opposite
 to that in the stiff system ($K_c=5.0$) as shown in Fig.\ref{fig:rate-JD}.
The convergence for $D$ means that
 the particle shows {\it normal} diffusion not only in equilibrium
 but also in non-equilibrium steady states.

If the particle steps stochastically,
 $J$ and $D$ should be related to the rate constants as follows.
\begin{equation}
|J| \simeq l (\kappa_F - \kappa_B), \quad
D \simeq \frac{l^2 (\kappa_F + \kappa_B)}{2},
\label{eqn:JD}
\end{equation}
 where $l$ is a lattice interval of the chain.
As is seen in Fig. \ref{fig:rate-JD}, these relations hold well
 especially in low temperature.
At high temperature, deviation from these relation occurs 
since the contribution from correlated successive stepping (multiple stepping)
 in small $\tau$ is not negligible.
Thus, it is confirmed that the stepping process is well characterized
 by the rate constants ($\kappa_F,\kappa_B$)
 at least within some lower temperature range.

\section{Bottleneck points as a rate-controlling factor}

For a simple chemical reaction, it is well known that rate constant $\kappa$
 obeys Arrhenius law 
$
\kappa\sim \exp(-{E_a}/{k_B T}),
$
where $E_a$ is called activation energy
 characterizing the reaction rate.
In transition state theory (TST), the activation energy is purely determined
 from the information of the potential energy surface along the reaction coordinate
 without considering temperature effect.
The minimum height of {\it energy barrier}, the difference of potential energy
 at the saddle point 
from that at the initial stable point
along the reaction coordinate,
 corresponds to the activation energy \cite{Hanggi}.

Far from equilibrium,
 we need to define which quantity can be used as the temperature
to characterize the process.
Because the particle position $x_h$ is the important observable for the stepping process,
the fluctuations of $\dot x_h$ in the staying state at a site would give the temperature.
This is approximately equal to kinetic temperature of the particle,
$T_h \equiv \left\langle {\dot x_h^2} \right\rangle$,
because the stepping event is rather rare and $J$ is sufficiently small
in the range of our analysis.
Here $\langle \cdot \rangle$ means ensemble and/or long-time average.
Under non-equilibrium condition ($T_2>T_1$),
$T_h$ takes values such that 
$T_2>T_h>T_1$
 due to the connection to the chain,
 while $T_h$ is equal to $T_2$ ($=T_1$) under equilibrium condition.

Let us consider the activation energy for the present system according to TST.
From the fixed point analysis for Eq.(\ref{eqn:model})\cite{Nakagawa_Kaneko5}, 
 the stable and the saddle points can be determined 
 so that the potential difference $\Delta U_0$ 
between them
 is obtained straightforwardly. 
Note that $\Delta U_0$ is a constant independent of temperature, while
depends on the flexibility of the system.
In the flexible system, there is a cooperative effect 
between the particle and the rail to make the activation energy lower. 
In the thermal equilibrium cases in Fig.\ref{fig:Arrhenius}, 
 the obtained $\Delta U_0$ actually works well as
 the activation energy in the Arrhenius law,
 characterizing the stepping rates 
 for the broad range of temperatures.
However, once the system is put under
 the non-equilibrium condition, 
 the rate constants deviate from the Arrhenius relation in equilibrium.

In the following, we try to get a modified version of
 the Arrhenius relation, applicable to non-equilibrium situation.
In TST, the saddle point is the most important
 for the calculation of the activation energy.
Although the usage of the saddle point itself might not be valid 
 far from equilibrium, the idea that
 some bottleneck point in phase space
 is the rate-controlling factor should not be necessarily abandoned.
How can we find such a ``bottleneck point''?
The paths for the stepping process is thermally fluctuating,
 but the ensemble of these paths would be recognized 
 as tube-like clouds in a phase space.
Then the most probable reaction path (MPRP) could be defined 
 by averaging over the tube-like clouds.
We suppose the bottleneck point would be defined
 as the saddle-like point along this MPRP.

In MPRP the particle position $x_h$, which is the major observable for the stepping process, would be the important coordinate in the phase space.
According to this idea, we propose that the ensemble which determines
 the bottleneck point consists of the points in the paths
 satisfying the conditions for the potential derivatives by $x_h$:
\begin{equation}
\frac{\partial V(x_h,\{x_i\})}{\partial x_h}=0, \quad
\frac{\partial^2 V(x_h,\{x_i\})}{\partial x_h^2}<0,
\label{eqn:neck}
\end{equation}
 with the supplemental condition that the points have been passed almost
 synchronously with the actual stepping events. 
 The condition (\protect\ref{eqn:neck}) is more gentle than
 that for the saddle points
 because it neglects the derivatives in other directions $\{x_i\}$.

The first condition of Eqs. (\ref{eqn:neck}) means that
 force acting on the particle is zero at the point and the second condition means
 that the point is an unstable fixed point if looking only along $x_h$.
By averaging the ensemble
 of the points satisfying the above condition with distinction
 of the direction of the stepping, bottleneck points for the forward
 and the backward stepping processes are obtained separately.

Interestingly, the bottleneck points for the forward and the backward stepping
 are often different in non-equilibrium ($T_2>T_1$),
 while the two points is always identical in equilibrium ($T_2=T_1$).
It suggests that the symmetry between the forward and
 the backward processes is violated in the presence of temperature difference. 

\section{Direction-dependent activation energies}

The difference of the bottleneck points will lead to the difference of
 the energy barrier. 
The barrier height $\Delta U_F$  ($\Delta U_B$) 
 can be determined for the forward (backward) stepping process 
 from the potential difference between the stable and 
the bottleneck point in similar manner to $\Delta U_0$.
It is noted that
 $\Delta U_F$ and $\Delta U_B$ could depend on the temperatures in contrast
 to $\Delta U_0$,
 because the bottleneck point is obtained as the ensemble average 
 of the points in the numerically simulated trajectories. 

In Figs.\ref{fig:U_neck}, $\Delta U_F$ and $\Delta U_B$ are 
 displayed as a function of the external driving (temperature difference).
In equilibrium ($T_1=T_2$), $\Delta U_F$ is equal to $\Delta U_B$ 
 as directly noticed from the identity of the bottleneck points
 between the two directions.
This is consistent with the equality of the rate constants $\kappa_R=\kappa_L$.
In non-equilibrium, the barrier heights varies from its equilibrium values,
 and their dependences on the external driving
 are different between the flexible and stiff system
as is seen in Figs. \ref{fig:U_neck}.

For the flexible system ($K_c=0.5$),
barrier heights grow with the increase of the external driving,
where the deviation of $\Delta U_F$
 from $\Delta U_0$ is smaller than that of $\Delta U_B$.
As clearly seen in Fig. \ref{fig:Arrhenius2},
 substituting $\Delta U_F$ and $\Delta U_B$
 for the activation energies,
 the rate constants in non-equilibrium can be reproduced
 by the Arrhenius-like form
\begin{equation}
\kappa_F = A \exp\left(-\frac{\Delta U_F}{k_B T_h}\right), \quad \kappa_B = A \exp\left(-\frac{\Delta U_B}{k_B T_h}\right),
\label{eqn:Arrhenius-like}
\end{equation}
 where $A$ is a temperature-independent prefactor
 approximately common in the forward and the backward stepping.
This relation indicates that the bottleneck points are
 major rate-controlling factors 
 and that $\Delta U_F$ ($\Delta U_B$) is
 the substantial {\it activation energy}
 for the forward (backward) stepping process far from equilibrium.
The knowledge of these activation energies
 is sufficient to characterize 
the oriented stepping process.

To be emphasized, 
 the direction-dependent activation energies 
mean that absolutely different paths are selected
 between the forward and the backward stepping reactions.
From the dependency of $\Delta U_F$ and $\Delta U_B$, 
 the reaction path for the backward stepping process is implied to be
more sensitive
 to the external driving than that for the forward. 
This sensitivity may be related to the property of the cooperative dynamics 
to realize the reaction path but this is open as a future problem.

For the stiff systems,
the fluctuations of the rail is small
and the cooperative dynamics between the particle and the rail
 are almost suppressed.
As a result, the bottleneck points do not draw apart from the saddle point
 with increasing external driving, and the energy barriers defined
 by these bottleneck points only slightly increase compared
 to $\Delta U_0$ (see Fig.\ref{fig:U_neck}(b)).
Such slight variation cannot account for the deviation of the rate constants
 in non-equilibrium seen in Fig.\ref{fig:Arrhenius}, so
 the barrier heights $\Delta U_F$ and $\Delta U_B$ does not explain
 the relative rate $\kappa_F/\kappa_B$.
Further investigation for the stiff systems is not considered here,
 but it can be done in parallel to the Brownian ratchet models
 which has no flexibility of the chain\cite{Sekimoto,Review-Ratchet}.

\section{Discussion}

In this Paper, we have discussed mechanisms to generate oriented process.
Far from equilibrium in the flexible system,
the direction-dependent activation energies (DAEs)
 account for the oriented process well.
For the appearance of DAEs due to dynamical coupling between components,
 the symmetry violation between forward and its reverse reaction paths
 is a key feature, and it never occurs in equilibrium.
Thus oriented processes due to DAEs
 cannot be explained from the equilibrium properties of the relevant systems.
Instead, we should develop a direct understanding of
 non-equilibrium phenomena. 
The fact that Arrhenius-like form could be recovered 
in non-equilibrium steady state might suggest such new direction
to find some robust structure to describe the system in non-equilibrium.

Direction-dependent activation energies are also obtained
for the stepping process in Kinesin biomotors
by analyzing the rate constants for the stepping motion 
in experimental sequences
\cite{Nishiyama_Higuchi,Taniguchi_Nishiyama}.
Comparing to the simple diffusion process without ATP-hydrolysis
exhibited by Kinesin mutants \cite{Okada_Hirokawa}
 (proposed as an equilibrium system),
the directionality in the normal functioning state
 seems to be generated by
 the suppression of the backward stepping process
 while leaving the rate of the forward process almost similar to that
in equilibrium \cite{Taniguchi_Nishiyama}.
Interestingly, our flexible system also exhibits a similar tendency
 under external driving.

{\bf Acknowledgment}

We are grateful to F.Oosawa, M.Nishiyama and Y.Taniguchi
for stimulating discussion to motivate this work;
M.Peyrard for fruitful discussion and critical reading of this manuscript;
and K.Kaneko for perpetual discussion from the beginning of this project.

\eject

\begin{figure}
\caption{
Profile of the model (refer to the text).
Potential form is asymmetric in space,
$
V(\Delta x)=  K_h{\tanh(p(\Delta x -r))}/{ \cosh(d\Delta x)},
$
where interaction is confined mostly to the nearest lattice sites 
($d^{-1}=0.25$ with $l=1$).
The degree of asymmetry is fixed at $p=10$ and $r=0.3$, while
coupling strength $K_h=0.2$. 
In numerical simulation, the chain consists of $7$ or $40$ lattice sites 
with periodic boundary condition, while $\gamma$ is fixed at $0.1$.
}
\label{fig:model}
\end{figure}

\begin{figure}
\caption{
Typical time sequence of $x_h$ 
for $K_c=0.5$, $T_1=0.02$, $T_2=0.08$ ($T_h=0.042$) and $T_1=T_2=0.042$. 
(Inset): $\tau$ vs $\bullet$ $\kappa_F(\tau)$, $\circ$ $\kappa_B(\tau)$ calculated according to the definition in Eq.(\ref{eqn:rate-eq}),
which show quick convergence. 
}
\label{fig:sequence}
\end{figure}

\begin{figure}
\caption{
Current $J$ and diffusion coefficient $D$ for $T_1=0.02$ with various values of $T_2$.
The guidelines are the relationship in Eq.(\ref{eqn:JD}).
$J$ changes its direction
from the flexible to the stiff system,
where it is positive in $x_h$ for $K_c=0.5$ while negative 
for $K_c=5.0$.
}
\label{fig:rate-JD}
\end{figure}

\begin{figure}
\caption{
Arrhenius plots, $1/T_h$ vs $\kappa_F, \kappa_B$.
In non-equilibrium (neq), we fix $T_1$ at $0.02$ while $T_2$ is widely changed,
i.e., the points in (neq) equal to those in (eq) at $1/T_h=50$.
The dotted lines are the Arrhenius law $\exp(-\Delta U_0/T_h)$,
where $\Delta U_0\simeq 0.15$ and $0.22$ for $K_c=0.5$ and $5.0$, respectively.
Prefactor is a fitted constant.
For $K_c=0.5$, $\kappa_F$(eq)$=\kappa_B$(eq)$\simeq\kappa_F$(neq)
while $\kappa_B$(neq)$<\kappa_B$(eq).
For $K_c=5.0$, both $\kappa_F$(neq)
and $\kappa_B$(neq) are suppressed.
}
\label{fig:Arrhenius}
\end{figure}

\begin{figure}
\caption{
The increase of the potential barrier $\Delta U_F$ and $\Delta U_B$.
The barrier height grows with $T_h-T_1$
for the flexible system but it does not for the stiff system. 
Upper ($Kc=0.5$): $T_1=0.02$ (circle), $T_1=0.01$ (rectangle).
Lower ($Kc=0.5$): $T_1=0.02$ (circle), $T_1=0.03$ (rectangle).
At the equilibrium point ($T_h=T_1$) for $K_c=0.5$, $\Delta U_F=\Delta U_B$ 
but there are slightly larger than $\Delta U_0$.
This deviation might be due to 'saddle point avoidance' \protect\cite{Berezhkovskii}. 
}
\label{fig:U_neck}
\end{figure}

\begin{figure}
\caption{
Exponential relationship between $\kappa_F (\kappa_B)$ and
$\Delta U_F/T_h (\Delta U_B/T_h)$ for $K_c=0.5$ ($k_B=1$).
Arrhenius-like relationship is recovered with regarding $\Delta U_B$ as the 
activation energy. 
}
\label{fig:Arrhenius2}
\end{figure}

\end{document}